\documentclass[fleqn,usenatbib]{mnras}

\usepackage{newtxtext,newtxmath}
\usepackage{amssymb, amsmath,framed}

\usepackage{bm}
\expandafter\ifx\csname package@font\endcsname\relax\else
 \expandafter\expandafter
 \expandafter\usepackage
 \expandafter\expandafter
 \expandafter{\csname package@font\endcsname}
\fi
\usepackage[T1]{fontenc}

\DeclareRobustCommand{\VAN}[3]{#2}
\let\VANthebibliography\thebibliography
\def\thebibliography{\DeclareRobustCommand{\VAN}[3]{##3}\VANthebibliography}

\usepackage{graphicx}	
\usepackage{amsmath}	
\usepackage{amssymb}	
\usepackage{graphicx}
\usepackage{bm}
\usepackage{color}

\title[TDEs and Galactic Habitability]{The Impact of Tidal Disruption Events on Galactic Habitability}

\author[Pacetti et al.]{
E. Pacetti$^{1}$, A. Balbi$^{1}$\thanks{E-mail: balbi@roma2.infn.it}, M. Lingam$^{2,3}$, F. Tombesi$^{1,4,5,6}$ and E. Perlman$^{2}$
\\
$^{1}$Department of Physics, University of Rome ``Tor Vergata'', Via della Ricerca Scientifica 1, I-00133 Rome, Italy\\
$^{2}$Department of Aerospace, Physics and Space Sciences, Florida Institute of Technology, Melbourne FL 32901, USA\\
$^{3}$Institute for Theory and Computation, Harvard University, Cambridge MA 02138, USA\\
$^{4}$INAF Astronomical Observatory of Rome, Via Frascati 33, I-00078 Monte Porzio Catone, Italy\\
$^{5}$Department of Astronomy, University of Maryland, College Park, MD 20742, USA\\
$^{6}$NASA - Goddard Space Flight Center, Code 662, Greenbelt, MD 20771, USA\\
}
\date{Accepted XXX. Received YYY; in original form ZZZ}

\pubyear{2020}

\begin{document}
\label{firstpage}
\pagerange{\pageref{firstpage}--\pageref{lastpage}}
\maketitle

\begin{abstract}
Tidal Disruption Events (TDEs) are characterized by the emission of a short burst of high-energy radiation. We analyze the cumulative impact of TDEs on galactic habitability using the Milky Way as a proxy. We show that X-rays and extreme ultraviolet (XUV) radiation emitted during TDEs can cause hydrodynamic escape and instigate biological damage. By taking the appropriate variables into consideration, such as the efficiency of atmospheric escape and distance from the Galactic center, we demonstrate that the impact of TDEs on galactic habitability is comparable to that of Active Galactic Nuclei. In particular, we show that planets within distances of $\sim 0.1$-$1$ kpc could lose Earth-like atmospheres over the age of the Earth, and that some of them might be subject to biological damage once every $\gtrsim 10^4$ yrs. We conclude by highlighting potential ramifications of TDEs and argue that they should be factored into future analyses of inner galactic habitability.
\end{abstract}




\section{Introduction}
The concept of habitability is complex due to its inherently multi-faceted nature. In particular, habitability is regulated by a plethora of processes spanning a diverse range of spatial and temporal scales. While most studies have exclusively focused on habitability at the planetary or stellar levels \citep{CBB16,LL19}, it has been increasingly appreciated over the past couple of decades that high-energy astrophysical phenomena may regulate habitability on galactic scales \citep{Gon05,Pran}. Most studies in this realm have tended to focus on the effects of supernovae and Gamma Ray Bursts; see the recent reviews by \citet{GM18,Kaib}.

Over the past few years, the impact of Active Galactic Nuclei (AGNs) has attracted greater attention, building on earlier studies \citep{Cla81,Gon05}. In particular, recent studies indicate that AGNs could drive the complete depletion of Earth-like atmospheres across a significant fraction of planets in galaxies \citep{BT17,FL18,WKB}, promote disruption of biospheres due to elevated radiation fluxes \citep{BT17,LGB,AC19}, and permit the synthesis of prebiotic compounds and carbon fixation \citep{LGB,LCD20}. A crucial property of AGNs should, however, be borne in mind: they are relatively short-lived, and only a small fraction of all galaxies are ``active'' at any given moment in time \citep{Krol99}.

Aside from AGNs, another crucial high-energy process intrinsically associated with supermassive black holes (SMBHs) is tidal disruption events (TDEs).\footnote{Although we will treat TDEs and AGNs as independent phenomena hereafter, \citet{PL20} recently suggested that the former might trigger the latter in some instances, especially at higher redshifts.} The existence of TDEs was predicted and modeled in the 1970s and 1980s \citep{Hil75,FR76,LTH82}. These phenomena arise when stars traverse too close to SMBHs, and are consequently disrupted by the latter's tidal field \citep{Kom15,Alex17,FWL20}. Despite the fact that TDEs are expected to recur in a wide range of galaxies, primarily those which host SMBHs of masses $\sim 10^6-10^8\,M_\odot$ \citep{SKC19}, there have been no studies devoted to assessing their impact on galactic habitability.

Hence, our goal is to explore the cumulative impact of TDEs by specifically focusing on the Milky Way and Sagittarius A* (Sgr A*) herein. The outline of this paper is as follows. In Sec. \ref{Method}, we describe some basic properties of TDEs and our methodology. We present the ensuing results in Sec. \ref{Results}, and analyze the ramifications for habitability in Sec. \ref{Conclusion}.

\section{Modeling the impact of TDEs on habitability}\label{Method}
In this Section, we describe our methodology to quantify the effects on habitability engendered by TDEs. Stars with mass $M_\star$ and radius $R_\star$ undergo tidal disruption if they approach the SMBH of mass $M_{BH}$ closer than the tidal radius $R_t \approx R_\star \left(M_{BH}/M_\star\right)^{1/3}$ \citep{Hil75}. During the course of a TDE, roughly one-half of the stellar mass escapes on a hyperbolic orbit, whereas the rest falls inward onto the black hole at a rate $\dot{M}$. The accretion rate is often modeled via the following power-law parametrization \citep{Rees,LFB15}:
\begin{equation}\label{AccRate}
    \dot{M} = \frac{M_\star}{3 t_{min}} \left(\frac{t+t_{min}}{t_{min}}\right)^{-5/3},
\end{equation}
where $t_{min}$ is defined as
\begin{equation}
    t_{min} = \frac{\pi}{\sqrt{2}} \left(\frac{R_t}{R_\star}\right)^{3/2} \sqrt{\frac{R_t^3}{G M_{BH}}} \approx 41\,M_6^{1/2}\,\mathrm{d},
\end{equation}
for a solar-type star, with $M_6 = M_{BH}/(10^6\,M_\odot)$. The power-law exponent of $-5/3$ in (\ref{AccRate}) is not accurate \emph{sensu stricto} \citep{AGR17}, but it suffices for our purposes. The energy acquired during accretion is re-emitted by the accretion disk over the span of $\sim 1$ yr \citep{Kom15} with a luminosity $L = \eta \dot{M} c^2$, where $\eta \sim 0.1$ is the typical radiative efficiency. By employing (\ref{AccRate}) in conjunction with the definition of $L$, the upper bound on the total energy ($E_{tot}$) emitted is given by
\begin{equation}\label{Energy}
  E_{tot} =  \int_{0}^\infty L\,dt = \frac{\eta M_\star c^2}{2},
\end{equation}
where the ``clock'' is started at $t = 0$ hereafter. In order to estimate the fraction $f_{XUV}$ of total energy emitted in the X-ray and extreme ultraviolet (XUV) band, we opt to model the TDE as a black body with a temperature of $k_B T_\star \approx 100$ eV \citep{SRK17}, and specify the XUV wavelength range of $1.24 \times 10^{-3} < \lambda < 1.24 \times 10^2$ nm, i.e., corresponding to energies of $10-10^6$ eV. For the black body ansatz, we find that $f_{XUV} \approx 0.99$, implying that the overwhelming majority of the total energy is emitted as XUV photons. 

\subsection{Hydrodynamic loss}\label{SSecHLoss}
The flux of XUV photons ($F_{XUV}$) at a given distance $D$ is
\begin{equation}\label{FXUV}
    F_{XUV} = \frac{f_{XUV} L e^{-\tau}}{4\pi D^2},
\end{equation}
under the assumption of isotropic radiation; note that $\tau$ denotes the effective optical depth along the line of sight that is discussed further below. The XUV incident on a planet at this distance is capable of driving atmospheric escape through a number of channels \citep{Ow19,LL19}. We will model the resultant atmospheric loss via energy-limited hydrodynamic escape \citep{BT17}, thereby yielding
\begin{equation}\label{Mlossdef}
    M_{lost} = \frac{3}{4} \frac{\varepsilon}{G \rho_p} \int_{0}^\infty F_{XUV}\,dt,
\end{equation}
where $M_{lost}$ is the mass lost during a single TDE, $\rho_p$ is the density of the planet, and $0.1 \leq \varepsilon \leq 0.6$ is the heating efficiency that describes the conversion of incident power into atmospheric loss. By combining (\ref{Energy}), (\ref{FXUV}) and (\ref{Mlossdef}), we obtain
\begin{equation}\label{Mloss}
  M_{lost} = \frac{3}{4} \frac{\varepsilon}{G \rho_p}  \frac{f_{XUV} e^{-\tau}}{4\pi D^2} \left(\frac{\eta M_\star c^2}{2}\right).
\end{equation}

There are two important caveats concerning our model. First, when it comes to $\tau$, there are two contributors: (i) the dusty torus at the center of an active galaxy \citep{UP95}, and (ii) the interstellar medium. While we take (i) into account along the lines described in \citet{BT17}, tackling (ii) is much more complex as it depends not only on the wavelength range but also $D$ and the galactic morphology (e.g., spiral or elliptical). Hence, as with prior publications on AGNs, we do not incorporate this contribution to $\tau$. The second limitation stems from the exclusion of the beaming effect in TDEs \citep{DMR18,FWL20}. Assessing the impact of beamed emission is rendered difficult, because the number of stars affected by the beamed emission depends not only on the solid angle $\Omega$ but also the orientation of the beam axis and the galactic morphology. For this reason, we restrict ourselves to studying isotropic emission hereafter.

\subsection{Biological hazards}\label{SSecBioH}
A proper assessment of the biological hazards entails more subtleties because the characteristics of putative biota are unknown, as are the properties of the planetary atmosphere under question. 

On account of these reasons, we will utilize a simplified approach to assess the impact of XUV radiation on biota. It is known that ionizing radiation becomes lethal to Earth-based organisms after a certain critical threshold is exceeded \citep{Dart11}. As laboratory experiments are performed over a finite time, the thresholds are typically expressed in units of fluences. We adopt critical fluences of $\mathcal{F}_c \approx 5 \times 10^5$ erg cm$^{-2}$ for eukaryotic multicellular lifeforms and $\mathcal{F}_c \approx 5 \times 10^7$ erg cm$^{-2}$ for prokaryotic microbes, based on \citet[Section 6.2]{SW02} and \citet{BT17}. 

The electromagnetic energy emitted in the XUV band during a single TDE ($E$) is found by integrating the luminosity over the characteristic timescale of $\Delta t \sim 1$ yr, analogous to estimating $E_{tot}$ in (\ref{Energy}), thus yielding
\begin{equation}
  E =  \int_{0}^{\Delta t} f_{XUV} \cdot L\,dt.
\end{equation}
For a solar-type star that undergoes tidal disruption, the above formula yields $E \approx 7 \times 10^{52}$ erg, which is consistent with prior theoretical estimates \citep[e.g.,][]{LK18}. Under the assumption of isotropic emission, the fluence $\mathcal{F}$ at distance $D$ is given by
\begin{equation}\label{Fluence}
    \mathcal{F} = \frac{E e^{-\tau}}{4\pi D^2} = 5.8 \times 10^8\,\mathrm{erg\,cm^{-2}}\,e^{-\tau}\,\left(\frac{D}{1\,\mathrm{kpc}}\right)^{-2}.
\end{equation}
In comparison, the high-energy electromagnetic radiation delivered by a supernova over the same timescale of $\sim 1$ yr at distance $D$ is roughly predicted to be \citep[Equation 1]{ES95}:
\begin{equation}\label{SNe}
    \mathcal{F} = 6.6 \times 10^7\,\mathrm{erg\,cm^{-2}}\,\left(\frac{D}{1\,\mathrm{pc}}\right)^{-2}.
\end{equation}
When the fluence becomes equal to the critical values delineated previously, the corresponding distances ($D_c$) are estimated using
\begin{equation}\label{D_C}
  D_c = \sqrt{\frac{E e^{-\tau}}{4\pi \mathcal{F}_c}}. 
\end{equation}

There are two caveats worth mentioning at this juncture. First, the threshold has been expressed in terms of fluence, but the timescale over which a given fluence is exceeded also plays a vital role.\footnote{To put it differently, even a small flux will eventually exceed $\mathcal{F}_c$ after sufficient time.} Over the span of $\sim 1$ yr considered above, the XUV fluence of the modern Sun is $\mathcal{F}_\oplus \approx 1.5 \times 10^8$ erg cm$^{-2}$, based on the XUV flux of $F_\oplus \approx 4.64$ erg cm$^{-2}$ s$^{-1}$ \citep{RGG05}. Hence, the XUV fluence of a Sun-like star is actually higher than (\ref{Fluence}) in the event that $D \gtrsim 1$ kpc. Second, the majority of XUV radiation will be absorbed by a thick atmosphere \citep{SW02,MT11}, implying that the above results are accurate only for highly tenuous atmospheres, and should therefore be viewed as upper bounds. However, regardless of the atmosphere, what is undeniable is that the top-of-atmosphere XUV fluence at $D \sim 0.1$ kpc is orders of magnitude higher than $\mathcal{F}_\oplus$ as well as a supernova at $\sim 1$ pc, as seen from (\ref{Fluence}) and (\ref{SNe}). In other words, the effects induced by TDEs at this distance would be comparable or higher than SNe at $\sim 1$ pc, which are expected to be considerable \citep{Gehrels2003,BW17}.

Hence, a different approach entails the adoption of (i) flux instead of fluence, and (ii) photon wavelengths that can penetrate through an Earth-like atmosphere. To this end, we adopt the methodology proposed in \citet{LGB}, that was based on \citet{MT11}. The basic idea is that doubling the top-of-atmosphere (TOA) UV-C flux, for a world akin to Hadean-Archean Earth, might result in large-scale extinction of biota. The flux in this wavelength range is determined via
\begin{equation}\label{FUVC}
    F_{UVC} = \frac{f_{UVC} L e^{-\tau}}{4\pi D^2},
\end{equation}
where $f_{UVC}$ is the energy fraction of radiation emitted at UV-C wavelengths. By employing the black body spectrum with $T_\star$, we obtain $f_{UVC}$. By equating $F_{UVC}$ with the TOA UV-C flux of $2.2 \times 10^3$ erg cm$^{-2}$ s$^{-1}$ \citep{RSKS}, we determine the value of $D$ at which disruption of the biosphere may be possible. We caution that, even when it comes to such worlds, it is well-known that $\lesssim 1$ m of water can effectively protect aquatic organisms from UV photons \citep{CM98}; similar shielding could arise from soils and UV screening compounds \citep{CK99}. Furthermore, terrestrial worlds with massive atmospheres (e.g., Super-Earths) or substantial haze densities are capable of conferring additional protection against XUV radiation \citep{ADM16,LL19}. It is, therefore, important to recognize that our estimates are partly heuristic.

\section{Results}\label{Results}
We apply the methodology outlined in the previous section and adumbrate the salient results.

\subsection{Effects on the atmosphere}\label{SSecAtm}
\begin{figure}
	\begin{center}
	    	\centering
		    \includegraphics[width=\linewidth]{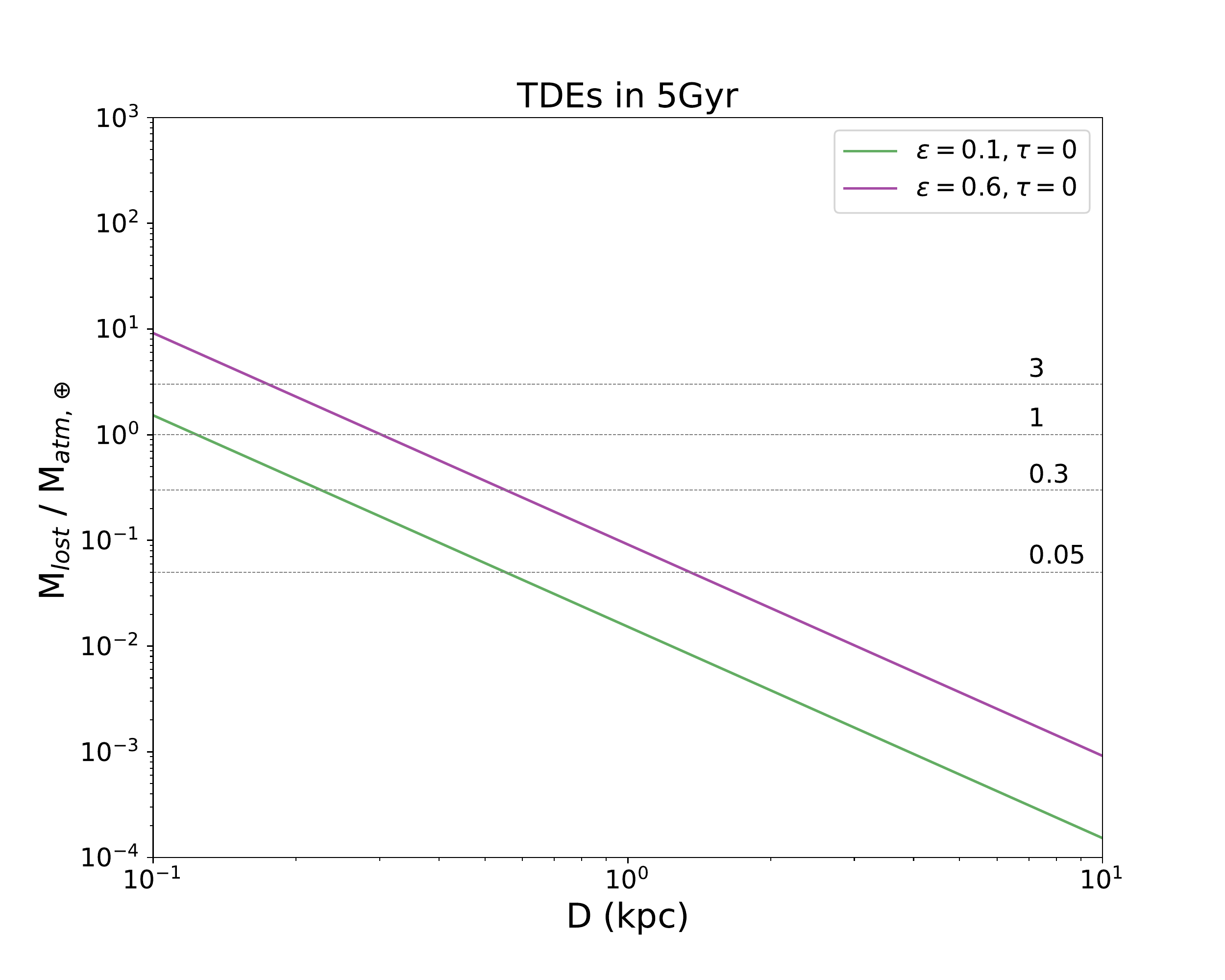}
		\end{center}
	\caption{Total mass (in units of Earth's present-day atmospheric mass) lost by a terrestrial planet at distance $D$ due to TDEs over a span of $5$ Gyr. We have adopted a TDE rate of $\sim 10^{-4}$ yr$^{-1}$ per galaxy. The results are shown for two values of the efficiency of hydrodynamic escape ($\varepsilon$). In both cases, we consider isotropic radiation propagating in an optically thin medium. }
		\label{Fig:TDE}
	\end{figure}

As one may expect, due to their transient nature, a single TDE does not cause significant atmospheric loss. A terrestrial planet at $D = 0.1$ kpc is depleted of merely $\sim 10^{-5}$ of the total mass of Earth's present-day atmosphere (denoted by $M_{atm,\,\oplus} \approx 5.1 \times 10^{21}$ g) for $\varepsilon = 0.6$ during a single TDE. Therefore, it is more instructive to calculate the cumulative atmospheric escape due to TDEs over a span of $5$ Gyr; this timescale is chosen as it corresponds to the age of our Solar system. We adopt a TDE rate of $\sim 10^{-4}$ yr$^{-1}$ per galaxy \citep{Kom15}. The results are depicted in Figure \ref{Fig:TDE}. 

\begin{figure*}
\begin{center}
			\centering
			\includegraphics[width=0.45\linewidth]{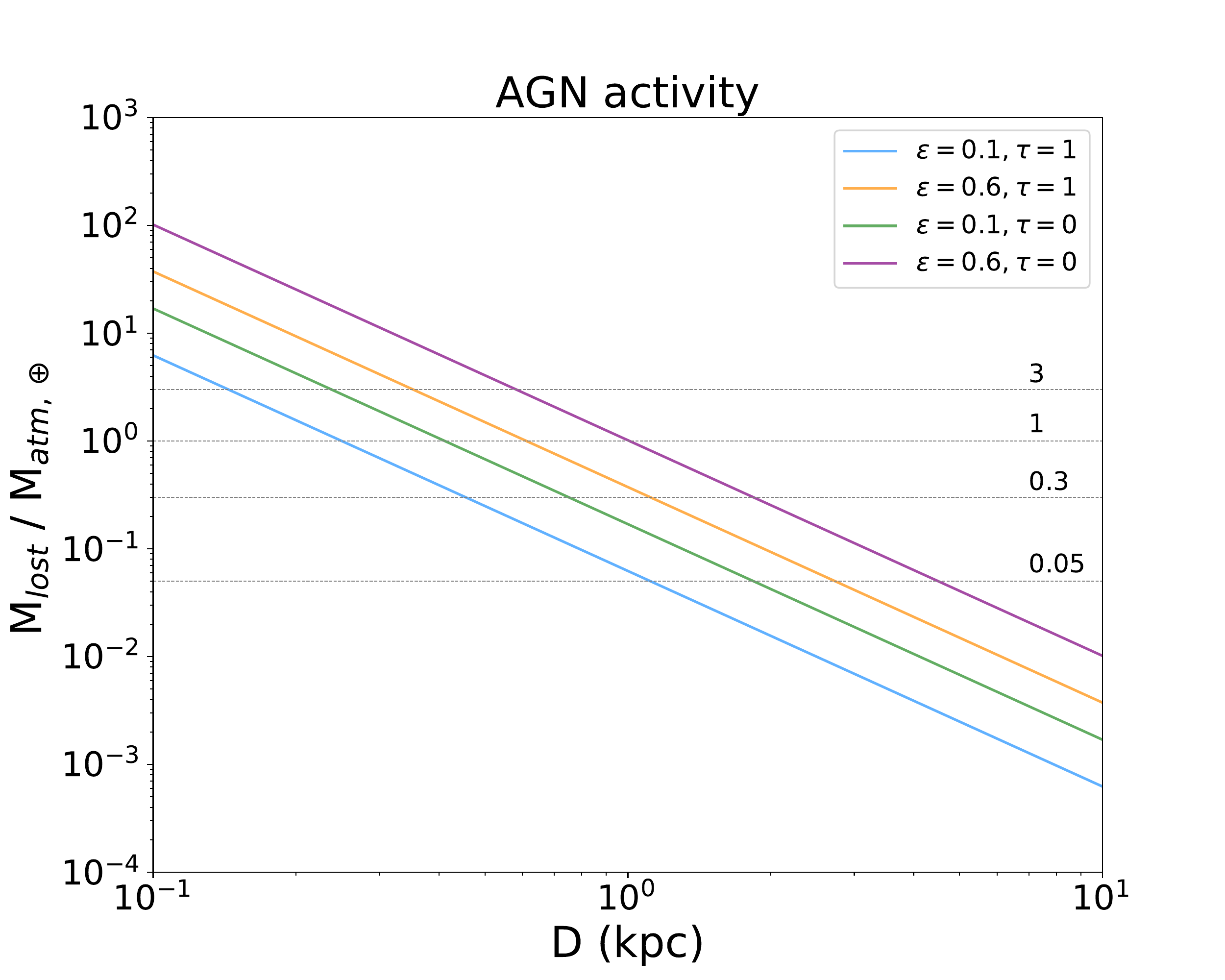}
			\includegraphics[width=0.45\linewidth]{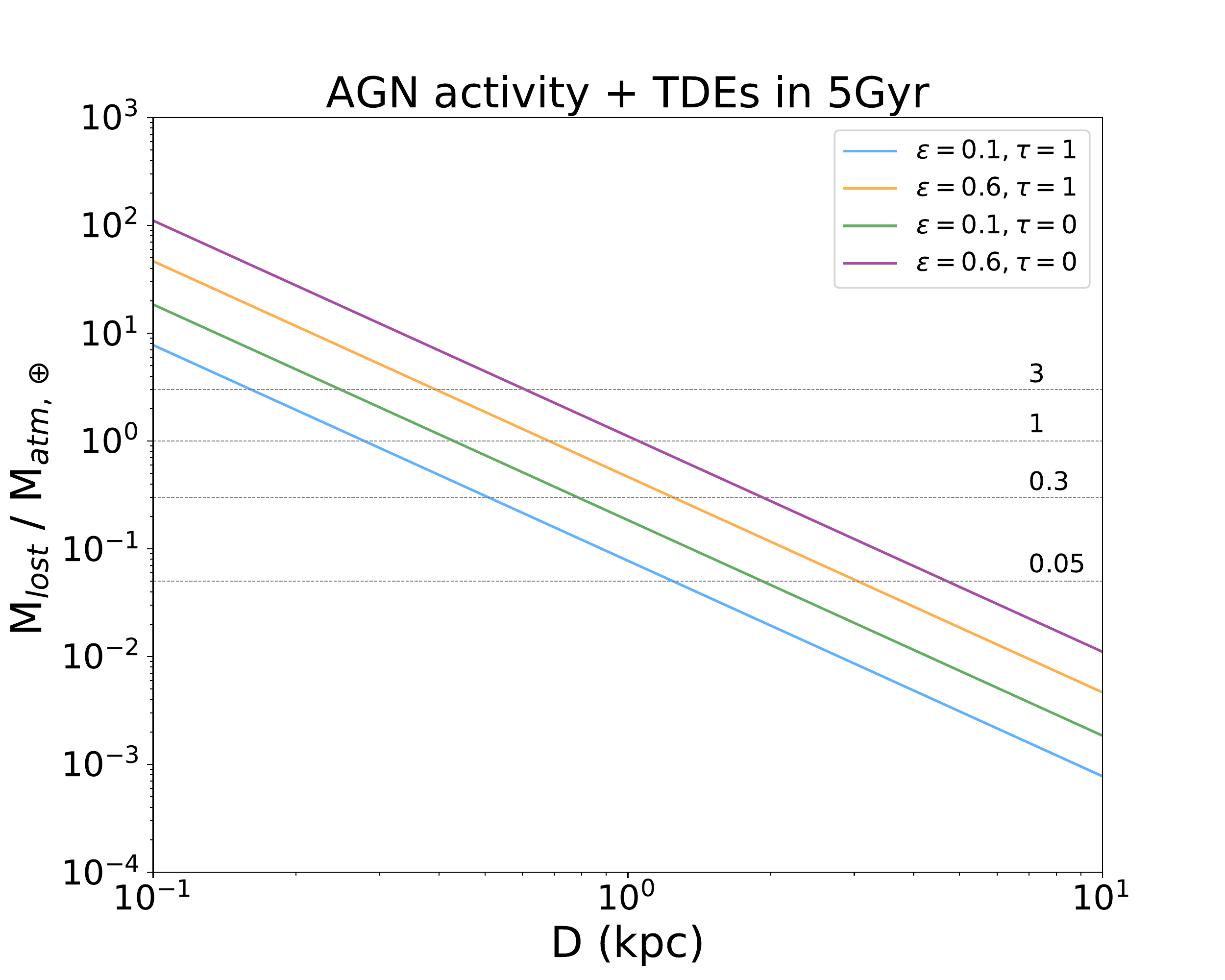}
\end{center}
\caption{Total mass (in units of the Earth'€™s present-day atmospheric mass) lost by a terrestrial planet at distance $D$. The left panel depicts the contribution from the AGN phase \citep{BT17}. The right panel illustrates the total radiative feedback from Sgr A*, i.e., considering the AGN phase and TDEs over a span of $5$ Gyr. We show the results for two values of the efficiency of hydrodynamic escape ($\varepsilon$) and the optical depth ($\tau$), which parametrizes the presence ($\tau=1$) or absence ($\tau=0$) of the obscuring torus during the AGN phase.}
\label{Fig:tdeagn}
\end{figure*}

It is important to recognize that the atmospheric loss from TDEs complements the depletion caused by the AGN phase. Hence, it is necessary to gauge the relative degree of atmospheric loss caused by TDEs and the AGN phase. In Figure \ref{Fig:tdeagn}, we show the atmospheric mass resulting from TDEs and the AGN phase. For fixed values of the free parameters, namely $D$, $\varepsilon$ and $\tau$, we observe that the mass loss due to TDEs is a few times smaller than the AGN phase. However, it is important to recall that many TDEs exhibit beamed emission \citep{DMR18}. When the beaming effect is included, the mass loss might become comparable to (or even higher than) the AGN phase, although uncertainties remain (see Sec. \ref{SSecHLoss}). On the other hand, the number of worlds that would be affected by beamed emission are correspondingly fewer than the isotropic case.

It is not easy to extract a generic result regarding the extent of the zone where atmospheric loss due to TDEs is prominent. The reason stems from the fact that there are three variables involved, namely, $\varepsilon$, $\tau$ and $\Omega$. Nonetheless, in all cases, we find that $M_{lost} \gtrsim M_{atm,\,\oplus}$ is fulfilled when the distance obeys $D\lesssim 0.1$-$1$ kpc. In contrast, if we are interested in the zone where Mars-like atmospheres are completely lost, it is more than an order of magnitude higher, i.e., the critical distance becomes $\sim 1$-$10$ kpc. On the other hand, the depletion of massive Super-Earth atmospheres is much more difficult, and thus shrinks the zone to $\sim 10$ pc.

Before moving ahead, a few caveats must be spelt out. The formalism in Sec. \ref{SSecHLoss} does not account for other source and sink terms \citep{CK17}, such as volcanic outgassing, continental weathering, and biological processes (e.g., photosynthesis). In other words, we assumed that the atmosphere exists in steady-state, with the sole exception of transient high-energy astrophysical processes that subsequently drive the intermittent atmospheric loss. Second, we invoked the paradigm of energy-limited hydrodynamic escape, but magnetohydrodynamic escape mechanisms are potentially dominant for certain inner Solar system worlds and exoplanets \citep{BBM16,DLMC,DJL18}.

Lastly, it is possible that the planet's atmosphere may return to its ``normal'' levels in the $\gtrsim 10^4$ yrs separating TDEs \citep{SVK20}. However, the loss of $10^{-5}$ of its atmosphere in a short timespan of 1 yr at the distance of $\sim 0.1$ kpc as noted earlier could nevertheless be detrimental. In particular, the escape rate becomes higher than its background value due to stellar irradiation by several orders of magnitude at $\sim 0.1$ kpc. Such a rapid increase has been posited as one of the chief reasons for the Triassic-Jurassic mass extinction \citep{WPZ14}, suggesting that similar detrimental effects are conceivable on other worlds.

\subsection{Biological damage}
We begin with estimating $D_c$ by employing (\ref{D_C}) for the two different values of $\mathcal{F}_c$ described in Sec. \ref{SSecBioH}. After further simplification, we obtain $D_c \approx 3.42$ kpc and $D_c \approx 34.2$ kpc for prokaryotes and eukaryotes, respectively, after using the parameters for Sgr A* and choosing $\tau = 0$. Therefore, only at distances $D > D_c$ would these organisms \emph{not} be susceptible to major damage. In comparison, \citet{BT17} carried out analogous calculations for the AGN phase of Sgr A*, and obtained $D_c \approx 1.33$ kpc and $D_c \approx 13.3$ kpc for prokaryotes and eukaryotes, respectively. Hence, insofar as $D_c$ is concerned, it would appear as though TDEs have a slightly more severe impact than the AGN phase.

At first glimpse, this appears to contradict the indubitable existence of complex muticellularity on Earth over the past $> 1$ Gyr \citep{Kno11}, because the Sun is $8$ kpc from the Galactic center, which is smaller than $D_c$. However, this is where the two caveats delineated in Sec. \ref{SSecBioH} come into play. To begin with, selecting $D = 8$ kpc in (\ref{Fluence}) yields a fluence that is more than an order of magnitude smaller than $\mathcal{F}_\oplus$, i.e., the solar XUV fluence in the same period. Moreover, the Earth's atmosphere would prevent the bulk of this XUV radiation from penetrating to the surface. Hence, we reiterate that the above two estimates for $D_c$ are upper bounds, and apply to worlds with very tenuous atmospheres.

The next factor that we consider is the biological damage via UV-C radiation incident on an Earth-analog, following the procedure described after (\ref{FUVC}). After selecting $\tau = 0$ and the parameters for Sgr A*, we obtain a critical distance of $0.6$ pc if we suppose that $L$ is the average luminosity of the TDE in (\ref{FUVC}), and $1.7$ pc if we use the peak luminosity instead. In comparison, the corresponding critical distance for Sgr A* during the active phase is $9.4$ pc \citep{LGB}, implying that the zones of biological damage associated with TDEs and AGNs are typically within an order of magnitude of each other.

However, the above distances are merely loose lower bounds on the zones of biological damage. At such close distances, the XUV fluence is many orders of magnitude higher than that from a Sun-like star, and will therefore have potentially severe consequences for biota, even when the majority is absorbed by the atmosphere. There is also the issue of atmospheric evaporation as noted in Sec. \ref{SSecAtm}; only Super-Earths with massive atmospheres would avoid total atmospheric loss at these distances \citep{CFL18}; such worlds are quite common in the Solar neighborhood \citep{Kal17}. Lastly, at such distances, planets are subjected to regular impacts by relativistic gas and dust that could pose additional impediments for habitability \citep{SDP18}. 

\section{Conclusion}\label{Conclusion}
In this paper, we analyzed the impact of TDEs on galactic habitability by taking two major factors dependent on high-energy radiation into consideration: atmospheric loss and biological damage. We chose to focus on Sgr A* because its fundamental parameters are well constrained, and due to the fact that we know of at least one planet hosting life (i.e., Earth) in the Galaxy. 

In the context of atmospheric depletion, we employed a simple hydrodynamic model to model this effect. We concluded that planets with Earth-like atmospheres may undergo total depletion over a span of $5$ Gyr up to distances of $D \sim 0.1$-$1$ kpc; the variation in the distance arises due to the free parameters involved. Next, we investigated the constraints on biological damage arising from TDEs. Although our model suggests that significant biological hazards are posed by the TDEs up to distances of $D \sim 10$ kpc, this constitutes a loose upper bound in all likelihood. At distances of $D \sim 0.1$-$1$ kpc, we found that the XUV fluences are higher than, or comparable to, those received by temperate planets around Sun-like stars, implying that worlds within this zone might experience non-trivial biological damage because of the higher radiation levels.

As per our findings, worlds at distances of $D \lesssim 0.1$-$1$ kpc may be susceptible to major perturbations of their biosphere by TDEs. Quite intriguingly, the mean interval between these disruptions is $\gtrsim 10^4$ yrs \citep{Kom15,SVK20}, which is not far removed from the characteristic timescales of $\sim 10^4-10^5$ yrs associated with Milankovitch cycles \citep{Berg88}. Therefore, for such worlds, it is conceivable that TDEs could potentially play the role of Milankovitch cycles in regulating climactic and biodiversity patterns. At even closer distances to the Galactic center, TDEs might initiate periodic mass extinctions, along the lines of Earth's alleged extinction periodicity of order $10$ Myr \citep{Bam06}. Whether life can persist over long timescales on this class of worlds would depend on the recovery timescale after extinction(s), which in turn is partly governed by the severity of the extinction(s); for instance, the most severe mass extinction in the Phanerozoic, in the end-Permian, entailed a recovery time of $\sim 10$ Myr \citep{CB12}.

To summarize, two broad conclusions emerge from this work. First, the cumulative deleterious impact of TDEs on habitability is broadly comparable to that of AGNs. Second, as the distance up to which the effects on surficial habitability are prominent could be $\sim 0.1$-$1$ kpc from the central black hole of the Milky Way, some fraction of the total number of planetary systems in the Milky Way within this region may have been adversely affected by the combined action of TDEs and the active phase of our Galaxy.\footnote{Note that these distances are applicable only in the context of \emph{surface-based} habitability. The number of worlds with subsurface oceans (and potentially with biospheres) might outnumber those with surface liquid water by a few orders of magnitude \citep{Ling19}.} Although there are some vital factors that have been set aside, our analysis suggests that TDEs might exert a substantive influence on planetary habitability at distances $\lesssim 0.1$-$1$ kpc from the central black hole of the MW. Hence, not only may they merit further investigation along the same lines as supernovae, gamma-ray bursts and AGNs, but they ought to be incorporated into state-of-the-art numerical models that track the spatio-temporal evolution of of galactic habitability at sub-kpc distances \citep{DCR15,FDC17,SVM19}.

\section*{Acknowledgements}
A.B.\ acknowledges support by the Italian Space Agency (ASI, DC-VUM-2017-034, grant number 2019-3 U.O Life in Space) and by grant number FQXi-MGA-1801 and FQXi-MGB-1924 from the Foundational Questions Institute and Fetzer Franklin Fund, a donor advised fund of Silicon Valley Community Foundation. 

\section*{Data availability}
The data underlying this article are available in the article.

\bibliographystyle{mnras}
\bibliography{TDE}

\bsp	
\label{lastpage}
\end{document}